\documentclass[arxiv]{melba}

\usepackage{amsmath,amsfonts,graphicx,hyperref} 
\usepackage{dblfloatfix}

\usepackage{amsmath,amsfonts}
\usepackage{afterpage}

\melbaid{YYYY:NNN}  
\doi{https://doi.org/10.59275/j.melba.2024-AAAA}
\melbaauthors{authors}  
\volume{2}
\firstpageno{1}  
\melbayear{2025}  
\datesubmitted{06/2025}  
\datepublished{07/2025}  

\melbaspecialissue{SurgiSR4K: A High‑Resolution Endoscopic Video Dataset for Robotic-Assisted Minimally Invasive Procedures}
\melbaspecialissueeditors{Marleen de Bruijne, Tal Arbel, Ismail Ben Ayed, Hervé Lombaert}

\ShortHeadings{SurgiSR4K: A High‑Resolution Endoscopic Video Dataset for Robotic-Assisted Minimally Invasive Procedures}{Fengyi Jiang and Xiaorui Zhang}

\title{SurgiSR4K: A High‑Resolution Endoscopic Video Dataset for Robotic-Assisted Minimally Invasive Procedures}

\author{
        \firstname Fengyi \surname Jiang\aff{1}\email{}\orcid{0009-0007-9420-5259}, 
	\firstname Xiaorui \surname Zhang\aff{1}\email{}\orcid{0009-0005-3639-3875},
	\firstname Lingbo \surname Jin\aff{1}\email{}\orcid{0000-0003-4735-8598},
        \firstname Ruixing \surname Liang\aff{1,2,3}\email{}\orcid{},
        \firstname Yuxin \surname Chen\aff{1,4}\email{}\orcid{0000-0003-3897-417X},
        \firstname Adi Chola \surname Venkatesh\aff{1}\email{}\orcid{0009-0003-5576-4876},
        \firstname Jason \surname Culman\aff{1}\email{},
        \firstname Tiantian \surname Wu\aff{5}\email{}\orcid{0009-0007-0198-6050},
        \firstname Lirong \surname Shao\email{}\aff{1},
        \firstname Wenqing \surname Sun\email{}\aff{1},
        \firstname Cong \surname Gao\email{}\aff{1}\orcid{0000x-0001-6798-8381},
        \firstname Hallie \surname McNamara\email{}\aff{1},
        \firstname Jingpei \surname Lu\email{}\aff{1}\orcid{0000-0002-9136-6096},
        \firstname Omid \surname Mohareri\email{}\aff{1}\orcid{}
}
\affiliations{
	\num 1 \addr Intuitive Surgical, Inc., Sunnyvale, CA, USA \\
        \num 2 \addr Johns Hopkins Medicine Neurosurgery, Baltimore, MD, USA \\
        \num 3 \addr Johns Hopkins University Electrical and Computer Engineering, Baltimore, MD, USA \\
        \num 4 \addr University of British Columbia Electrical and Computer Engineering, Vancouver, BC, Canada \\
        \num 5 \addr Wilford \& Kate Bailey Small Animal Teaching Hospital, Auburn, AL, USA \\
}

\abstract{%
High-resolution imaging is crucial for enhancing visual clarity and enabling precise computer-assisted guidance in minimally invasive surgery (MIS). Despite the increasing adoption of 4K endoscopic systems, there remains a significant gap in publicly available native 4K datasets tailored specifically for robotic-assisted MIS. We introduce SurgiSR4K, the first publicly accessible surgical imaging and video dataset captured at a native 4K resolution, representing realistic conditions of robotic-assisted procedures. SurgiSR4K comprises diverse visual scenarios including specular reflections, tool occlusions, bleeding, and soft tissue deformations, meticulously designed to reflect common challenges faced during laparoscopic and robotic surgeries. This dataset opens up possibilities for a broad range of computer vision tasks that might benefit from high resolution data, such as super resolution (SR), smoke removal, surgical instrument detection, 3D tissue reconstruction, monocular depth estimation, instance segmentation, novel view synthesis, and vision-language model (VLM) development. SurgiSR4K provides a robust foundation for advancing research in high-resolution surgical imaging and fosters the development of intelligent imaging technologies aimed at enhancing performance, safety, and usability in image-guided robotic surgeries.}


\begin{document}
\twocolumn[{%
  \maketitle  
}]


\section{Background}
    \enluminure{M}{inimally} invasive surgery (MIS), including laparoscopic and robot-assisted procedures, has transformed surgical practice by reducing patient recovery time, minimizing tissue trauma, and improving cosmetic outcomes \citep{mack2001minimally}. As these techniques become the standard of care in a wide range of clinical specialties, from general surgery to gynecology and urology, the demand for enhanced surgical vision and intra-operative guidance has increased in parallel \citep{conrad2015role, oberlin2016effect}. In particular, the quality of intra-operative imaging has become increasingly vital for ensuring surgical safety, precise dissection, and the detection of critical anatomical features such as vessel bifurcations, lesion margins, or hidden pathology \citep{marshall2010near, goh2010minimally}. For example, accurate margin assessment during colorectal or liver surgery can directly impact long-term patient outcomes \citep{montalti2015impact}.

\begin{figure*}[t]
    \centering
    \includegraphics[width=0.8\textwidth]{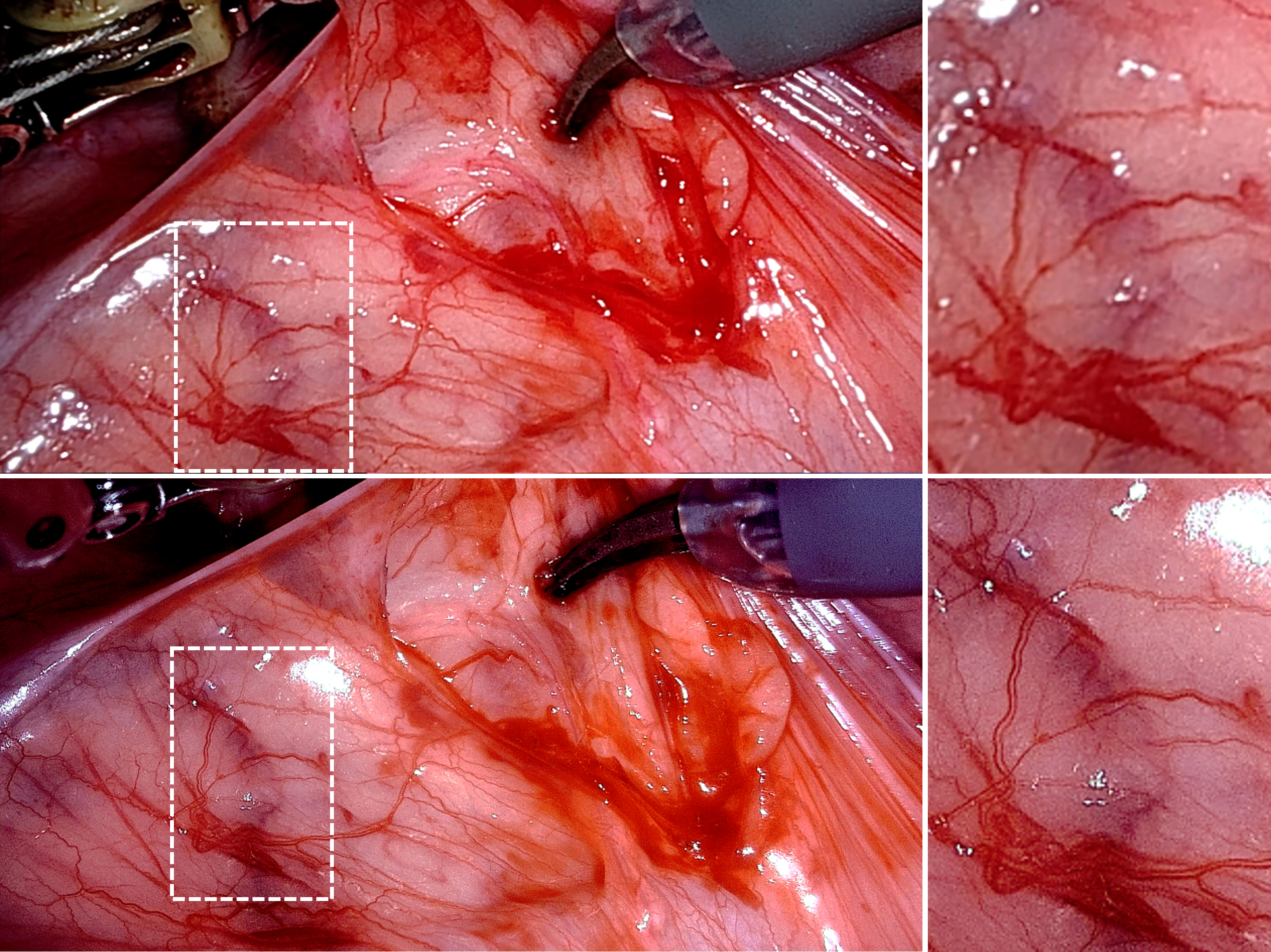}
    \caption{Side-by-side comparison of 1080p (top) and 4K (bottom) endoscopic images captured simultaneously with separate 1080p and 4K cameras.}
    \label{fig:resolution_comparison}
\end{figure*}
    
    In recent years, advances in medical imaging have driven a shift toward higher-resolution acquisition systems. In the domain of MIS, modern laparoscopy platforms such as Olympus VISERA or Stryker 1688 have begun offering ultra-high-definition (UHD) 4K images, promising substantial benefits in surgical navigation and decision making \citep{abdelrahman2018acquiring}. High resolution visualization enables surgeons to better differentiate tissue types, observe microvasculature, and assess resection margins, capabilities that are crucial for both oncologic and reconstructive interventions, as shown in Figure \ref{fig:resolution_comparison}. Recent clinical and simulator studies demonstrate that 4k laparoscopy shortens the operative time and reduces blood loss compared to HD systems \citep{puccetti2023}. Randomized skill acquisition trials also show faster learning curves and fewer errors for 4K vision \citep{abdelrahman2018acquiring}, and systematic evidence confirms that many performance metrics are close to those of dedicated 3D systems when 4K is used \citep{harada2018}.
     In addition, detailed imaging facilitates the development of downstream computer-assisted intervention (CAI) tools, such as automated segmentation, tissue classification, and image-based depth estimation. 4K resolution is widely available in non-surgical commercial space, yet most public surgical vision instruments remain trapped in the HD/sub-HD era, preventing algorithms from unleashing their full potential on utilizing high-resolution images.
    
    \begin{figure*}[b]
		\centering
		\includegraphics[width=1.0\linewidth]{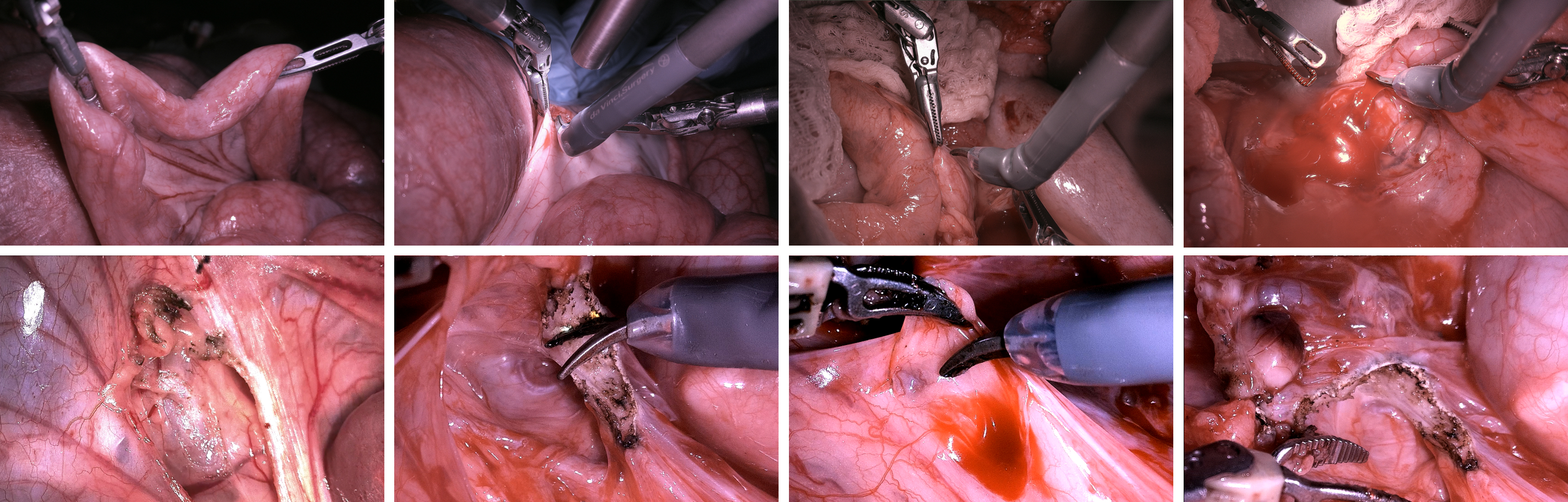}
		\caption{Example frames from the training dataset, showcasing various tools used in different scenarios. These frames highlight the diversity of situations included in the dataset.}
        \label{fig: samples images from dataset}
    \end{figure*}

\begin{table*}[ht]
\centering
\resizebox{\textwidth}{!}{%
\begin{tabular}{lccccc}
    \toprule
    \textbf{Dataset (Year)} & \textbf{Modality} & \textbf{Size} & \textbf{Native Res.} & \textbf{Primary Tasks} & \textbf{Licence / Access} \\
    \midrule
    ETIS-LaribPolyp \citep{silva2014toward}       & Colonoscopy stills           & 196 imgs              & 1225×966p      & polyp seg.          & free-research \\
    CVC-ClinicDB \citep{bernal2015wm}             & Colonoscopy stills           & 612 imgs              & 384×288p       & polyp seg.          & academic-only \\
    Cholec80 \citep{twinanda2016endonet}          & Laproscopic video            & 80 vids               & 854×480p       & cls., det, seg      & CC-BY 4.0 \\
    Kvasir \citep{pogorelov2017kvasir}            & Gastroscopy stills           & 8\,000 imgs           & 720p           & cls., det.          & CC-BY 4.0 \\
    Kvasir-SEG \citep{jha2019kvasir}              & Gastroscopy stills           & 1\,000 imgs           & 1000×1000p     & polyp seg.          & CC-BY 4.0 \\
    HyperKvasir \citep{borgli2020hyperkvasir}     & Gastroscopy video/stills     & 110\,k imgs, 374 vids & 720–1080p      & cls., seg., QA      & CC-BY 4.0 \\
    GIANA \citep{guo2019giana}                    & Colonoscopy imgs/videos      & 300 imgs, 30 vids     & 384–1080p      & det., seg.          & challenge-EULA \\
    EndoVis 2017 \citep{allan20192017}            & Robotic surgery video        & 15 seqs.              & 1920×1080p     & tool seg., track    & request-form \\
    EndoVis 2018 \citep{allan20202018}            & Robotic surgery video        & 15 seqs.              & 1280×1024p     & tool seg., track    & request-form \\
    HeiSURF \citep{wagner2023comparative}         & Cholecystectomy video        & 33 vids               & 720–1080p      & cls., det           & CC-BY 4.0 \\
    SurgVU \citep{zia2025surgical}                & Robotic surgery video        & 280 vids              & 1280×720p      & seg, det, cls       & CC-BY 4.0 \\
    \textbf{SurgiSR4K (2025)}\textsuperscript{*}  & Robotic surgery imgs/videos  & \textbf{800 imgs, 50 vids} & \textbf{3840×2160p} & \textbf{SR, seg, det} & \textbf{CC-BY-NC-SA 4.0/request-form} \\
    \bottomrule
\end{tabular}

}
\vspace{0.5em}
\caption{Comparison of public endoscopic and surgical datasets. \textbf{SurgiSR4K} is our proposed dataset introduced in this work.}
\label{tab:datasets}
\end{table*}

	

\subsection{Related Works}
Despite the growing availability of 4K endoscopic systems, the integration of 4K imaging into routine surgical practice and robotic-assisted practice remains limited due to high hardware costs, bandwidth constraints, and legacy infrastructure. Compounding this challenge is the surprising lack of publicly available 4K surgical datasets particularly those designed to support super-resolution (SR) and AI-driven computer-assisted intervention research. Early deep SR frameworks such as EndoL2H demonstrated clinically perceptible gains in capsule footage \citep{Almalioglu2020}. More recent transformer architectures further boost video PSNR in operative scenes \citep{Song2022,Zhang2023}, but are still constrained by sub-4K sources. SurgiSR4K provides the \textbf{first} native-resolution benchmark to close this training–deployment mismatch.
\begin{figure*}[!t]
		\centering
		\includegraphics[width=1.0\linewidth]{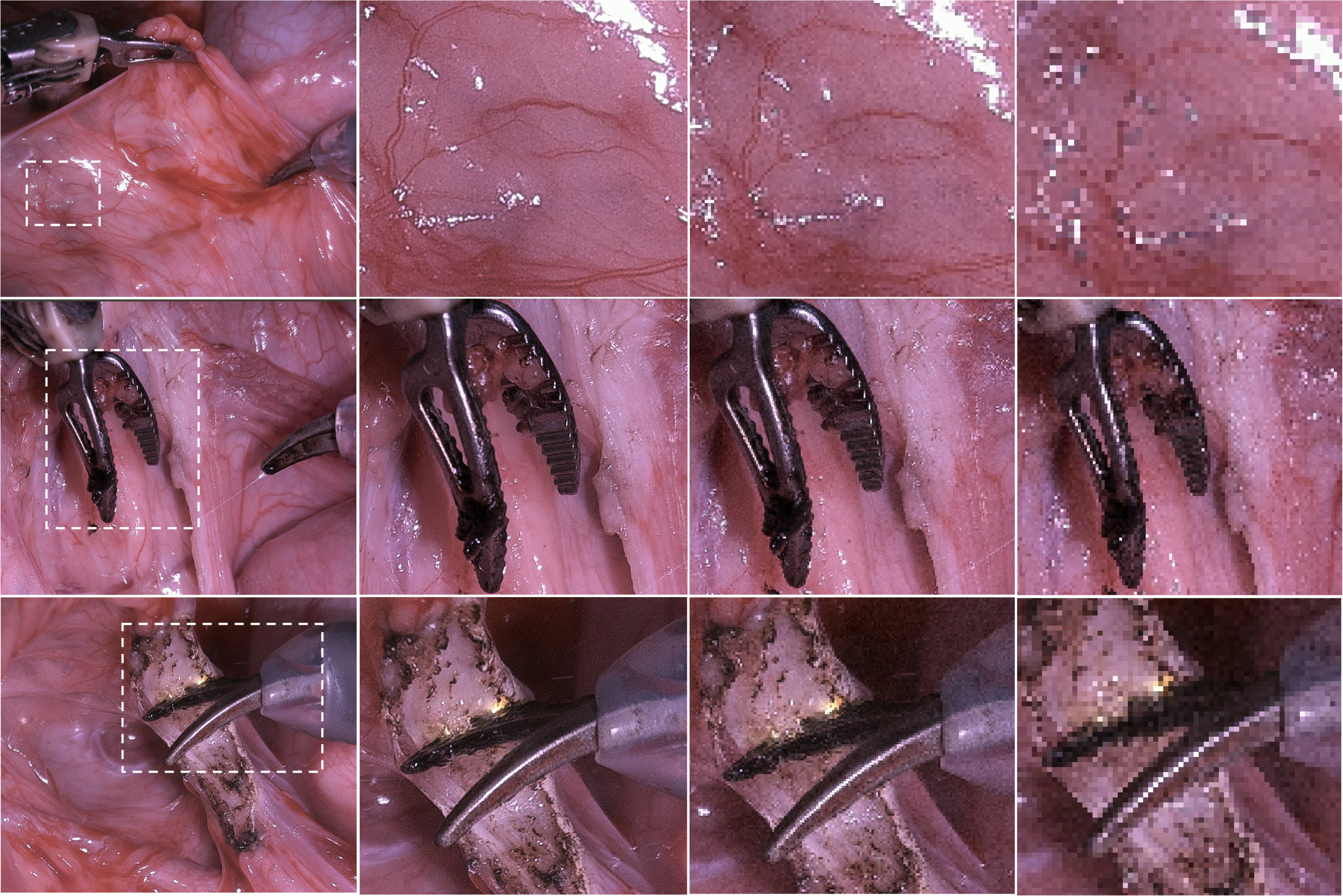}
		\caption{Comparison of image quality at different resolutions. From left to right: native 4K image (3840×2160), zoomed-in region of the 4K image, zoomed-in region of the downsampled 540p image (960×540), and zoomed-in region of the downsampled 270p image (480×270).}
        \label{fig: res compare}
\end{figure*}

The benefits of high-resolution imaging extend beyond SR, offering significant advantages for tasks such as monoscopic depth estimation and novel view synthesis \citep{xu20244k4d}. Unlike stereo endoscopes, which require dual-camera calibration and increase hardware complexity, monoscopic systems can potentially infer depth from a single high-resolution input. This is particularly valuable in robotic platforms where camera movement is restricted or in laparoscopic systems that use fixed-angle scopes. When trained on 4K video, models can better extract fine spatial details and textural cues necessary for robust depth estimation. These depth maps, in turn, support consistent novel view synthesis, enabling virtual camera repositioning, depth-aware overlays, or immersive visualization in education and planning.\citep{10782148}

    As shown in Table~\ref{tab:datasets}, currently available surgical vision datasets are limited in resolution, with most capped at 1280x1024 pixels \citep{allan20202018}, 1350×1080 pixels for odometry-focused datasets such as EndoSLAM \citep{Fredrick2022} or 1920×1080 pixels \citep{allan20192017}. Although these datasets have enabled progress in many vision tasks, their resolution is insufficient for training high-performance models that rely on capturing fine anatomical and contextual details. To address this critical gap, we introduce SurgiSR4K, the first high-quality, 4K-resolution surgical image and video dataset dedicated to enabling super-resolution and image understanding research in the context of minimally invasive procedures.

    
    To our knowledge, SurgiSR4K represents the first dataset of its kind designed explicitly for applications that require high-resolution data in surgical settings.  We anticipate that it will play a pivotal role in bridging the gap between state-of-the-art vision models and the operational needs of computer-assisted surgical platforms. As surgical AI continues to move from proof-of-concept to operating room integration, data sets such as SurgiSR4K will be indispensable in training, validating, and deploying reliable visual intelligence systems.
    
\section{Summary}
    SurgiSR4K represents a significant advancement in surgical imaging datasets, providing the first publicly available collection of native 4K resolution endoscopic images and videos specifically designed for robotic-assisted minimally invasive surgery research. The dataset comprises 800 high-quality 4K PNG images and 50 video clips (each 5 seconds at 30 fps) (\autoref{tab:dataset_composition}) captured from porcine animal models under realistic surgical conditions using a Da Vinci Xi surgical robot system.
    
    Unlike existing surgical datasets that are limited to HD or sub-HD resolutions, SurgiSR4K offers unprecedented visual detail at 3840×2160 pixels, enabling researchers to explore the full potential of high-resolution imaging in surgical AI applications. The dataset includes diverse surgical scenarios featuring various robotic instruments (monopolar curved scissors, force bipolar, bipolar forceps, grasping retractor, and Cadiere forceps) and challenging visual conditions such as specular reflections, tool occlusions, bleeding, smoke dispersion, and soft tissue deformations.
    
    To support multiscale research, each 4K image and video frame is accompanied by downsampled versions at 960x540p and 480x270p resolutions using Lanczos resampling (Fig. \ref{fig: res compare}), creating perfectly aligned multi-resolution triplets ideal for super-resolution training and evaluation. The dataset is structured to facilitate a broad range of computer vision tasks, including superresolution, monoscopic depth estimation, surgical instrument detection and tracking, instance segmentation, 3D tissue reconstruction, and novel view synthesis.
    
    All content has been clinically validated by experienced medical professionals and is released under the CC-BY-NC-SA 4.0 license to promote responsible use of research. SurgiSR4K addresses a critical gap in surgical AI research by providing the highest resolution surgical imaging data set available to date, enabling the development of next-generation computer-assisted surgical systems that can leverage the full benefits of 4K endoscopic imaging technology.

\begin{figure*}[!t]
		\centering
		\includegraphics[width=1.0\linewidth]{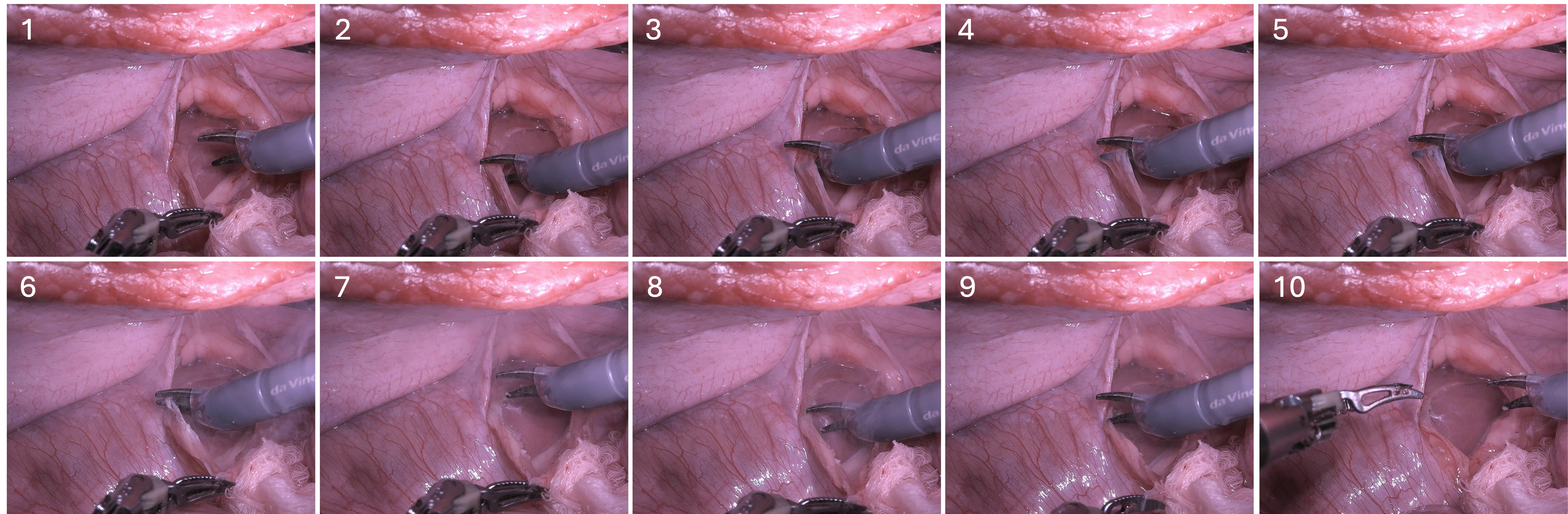}
		\caption{Frame-by-frame example of a 5-second video clip sampled at 2 fps, showing training data with a bipolar forceps instrument in a cauterizing motion.}
        \label{fig:video clip}
\end{figure*}

\section{Resource Availability}

\subsection{Potential Use Cases}
Super-resolution is a critical enabler for enhancing legacy or bandwidth-limited surgical video streams, particularly in low-cost or resource-constrained settings. SurgiSR4K provides a unique benchmark for SR algorithms by offering perfectly aligned multi-resolution image triplets (270p, 540p, and native 4K, as shown in Figure \ref{fig: res compare}). These triplets allow for supervised training of state-of-the-art models (e.g., SwinIR \citep{liang2021swinir}, Real-ESRGAN \citep{wang2021real}) as well as quantitative evaluation using perceptual and fidelity metrics. 

\subsection{Data Location}
The SurgiSR4K dataset—including video clips and corresponding high-resolution images—is publicly available through our dedicated repository on \href{https://www.synapse.org/Synapse:syn68756003}{\textcolor{blue}{\underline{Synapse:syn68756003}}}. The portal provides open access to the complete dataset, along with detailed metadata, project wiki, and supporting documentation.

\subsection{Licensing}
The SurgiSR4K dataset is released under the Creative Commons Attribution-NonCommercial-ShareAlike 4.0 International (CC BY-NC-SA 4.0) license. This permits users to share and adapt the dataset for non-commercial purposes, provided appropriate credit is given, any changes are indicated, and derivative works are distributed under the same license. 

\subsection{Ethical Considerations}
This animal model dataset was collected in vivo from healthy Yorkshire swine under general anesthesia, with no human subjects involved.  

All procedures were approved and conducted in accordance with Intuitive Surgical’s Institutional Animal Care and Use Committee (IACUC) protocol No. IACUC‑2022‑1200 (Title: [004] Use of Porcine Models for Robotic Surgical and Procedure Training). The study was conducted in alignment with the 3Rs principles—Refinement, Reduction, and Replacement—to ensure the ethical and humane treatment of animals.

All procedures were supervised by qualified veterinary staff, and personnel involved were trained and approved under the relevant IACUC protocols.

\begin{table}[h]
\centering
\resizebox{\columnwidth}{!}{%
\begin{tabular}{lc}
\toprule
\textbf{Dataset Component} & \textbf{Specification} \\
\midrule
\textbf{Native Resolution} & 3840×2160 (4K) \\
\textbf{Downsampled Resolutions} & 960×540 (540p), 480×270 (270p) \\
\textbf{Resampling Method} & Lanczos \\
\midrule
\textbf{Training Set} & \\
\quad Images & 800 PNG files \\
\quad Video clips & 50 AVI files \\
\quad Video duration & 5 seconds each \\
\quad Frame rate & 30 fps \\
\midrule
\textbf{Test Set (Hidden)} & \\
\quad Images & 300 PNG files \\
\quad Video clips & 50 AVI files \\
\midrule
\textbf{Surgical Instruments} & \\
\quad Monopolar curved scissors & \checkmark \\
\quad Force bipolar & \checkmark \\
\quad Bipolar forceps & \checkmark \\
\quad Grasping retractor & \checkmark \\
\quad Cadiere forceps & \checkmark \\
\bottomrule
\end{tabular}%
}
\caption{SurgiSR4K dataset composition and specifications.}
\label{tab:dataset_composition}
\end{table}

\section{Methods}            
\subsection{Experimental Setup}

\begin{figure}[!t]
  \centering
  \includegraphics[width=1\linewidth]{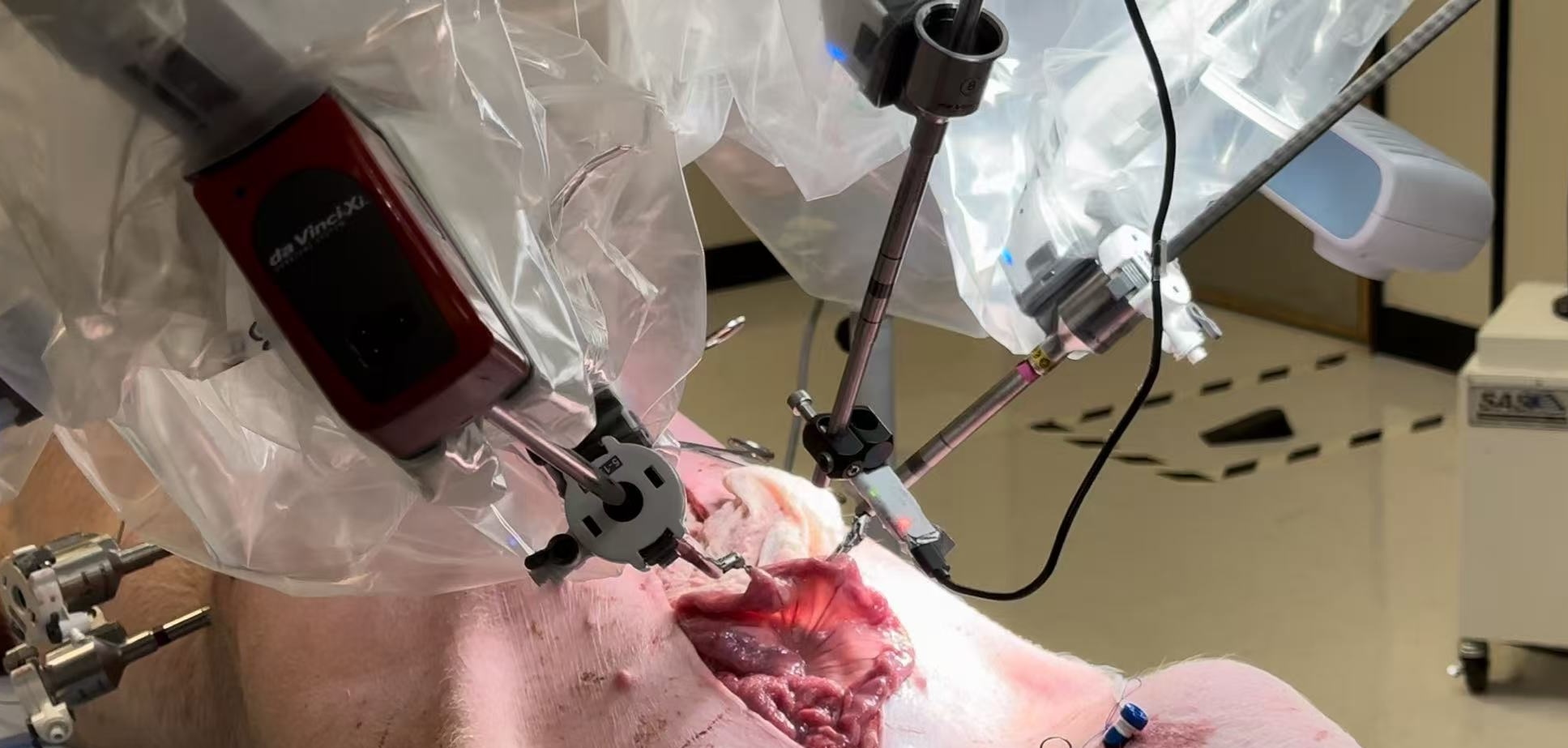}
  \caption{Experimental setup for SurgiSR4K data collection. Live porcine surgeries were performed using a Da Vinci Xi robot.}
  \label{fig:setup}
\end{figure}

To construct the SurgiSR4K dataset, we performed a series of open-scene porcine lab surgeries using endoscopic imaging, designed to emulate realistic minimally invasive surgical conditions. These procedures were performed by experienced veterinarians on live anesthetized swines, using an open abdominal and thoracic approach to allow direct access for native 4K external camera capture. A Da Vinci Xi surgical robot performed interventions on internal organs during each session, ensuring authentic instrument motion and tissue interaction, as shown in Figure \ref{fig:setup}.  Each recorded sequence spans several minutes and features diverse surgical events, including tool usage, organ manipulation, cautery smoke, bleeding, and tissue ablation, providing a rich source of visual complexity for downstream computer vision tasks (see Figure \ref{fig:video clip}). 

\paragraph{Equipment}
The imaging setup used a Sony IMX412 external camera module,  manufactured by Sony Corporation, featuring a 12.3 MP CMOS sensor (Sony Starvis series), with an optical stack of 74.4 ° diagonal (60.2 ° horizontal). \paragraph{Acquisition Settings}
Images and videos were recorded at a native 4K Ultra HD resolution of 3840×2160 pixels at 60 frames per second (fps). The camera was interfaced through CSI-2 to an NVIDIA Jetson TX2 Development Kit for real-time data handling and storage.
\paragraph{Subject Information}
\begin{itemize}
    \item Demographics: The dataset comprises adult pigs used under controlled experimental conditions. The animals were healthy and standardized by weight, anesthetized following institutional animal care guidelines.

    \item Cohort Description: The selection criteria ensured that the subjects were free of systemic diseases or anatomical anomalies that could affect surgical simulations. Multiple wet-lab procedures were performed to maximize the diversity of captured surgical scenarios.
\end{itemize}

\subsection{Dataset Composition}
\begin{figure}[!t]
  \centering
  \includegraphics[width=1.0\linewidth]{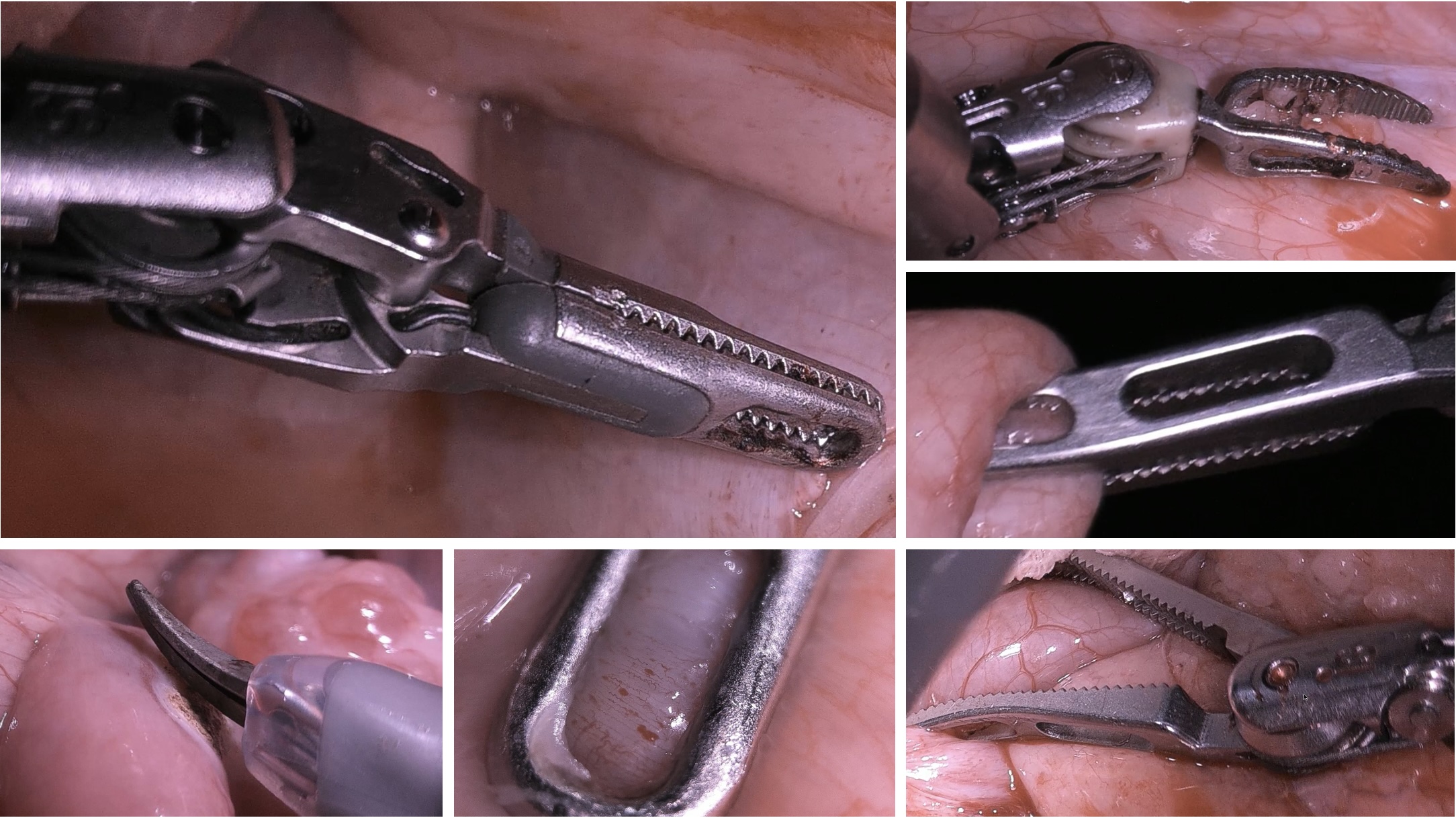}
  \caption{Example of various instruments included in the dataset, showcasing a range of tools used in different surgical or procedural contexts.}
  \label{fig:tools}
\end{figure}

Images and videos were captured at native 4K resolution (3840×2160), then downsampled to 540p (960×540) and 270p (480×270) using Lanczos resampling to support a range of super resolution tasks and multiscale training. The curated dataset includes 800 high-quality PNG images and 50 short video clips (AVI format), each 5 seconds in duration at 30 fps, selected for their diversity and visual richness (see Figure \ref{fig: samples images from dataset}, \ref{fig:video clip}). The data set features a variety of minimally invasive robotic surgical instruments, including monopolar curved scissors, bipolar forceps, grasping retractor, and cadiere forceps, improving its utility for tool recognition and interaction modeling tasks, as shown in Figure \ref{fig:tools}. An additional 300 images and 50 video clips have been reserved as a hidden test set for benchmarking and evaluation purposes.

\section{Validation}
All images and videos incorporated in the SurgiSR4K dataset were individually reviewed by an experienced Clinical Development Engineer (CDE) and an experienced veterinarian (VMD) to ensure clinical relevance and visual clarity for downstream applications. 

\section{Discussion}
\begin{figure*}[!t]
		\centering
		\includegraphics[width=1.0\linewidth]{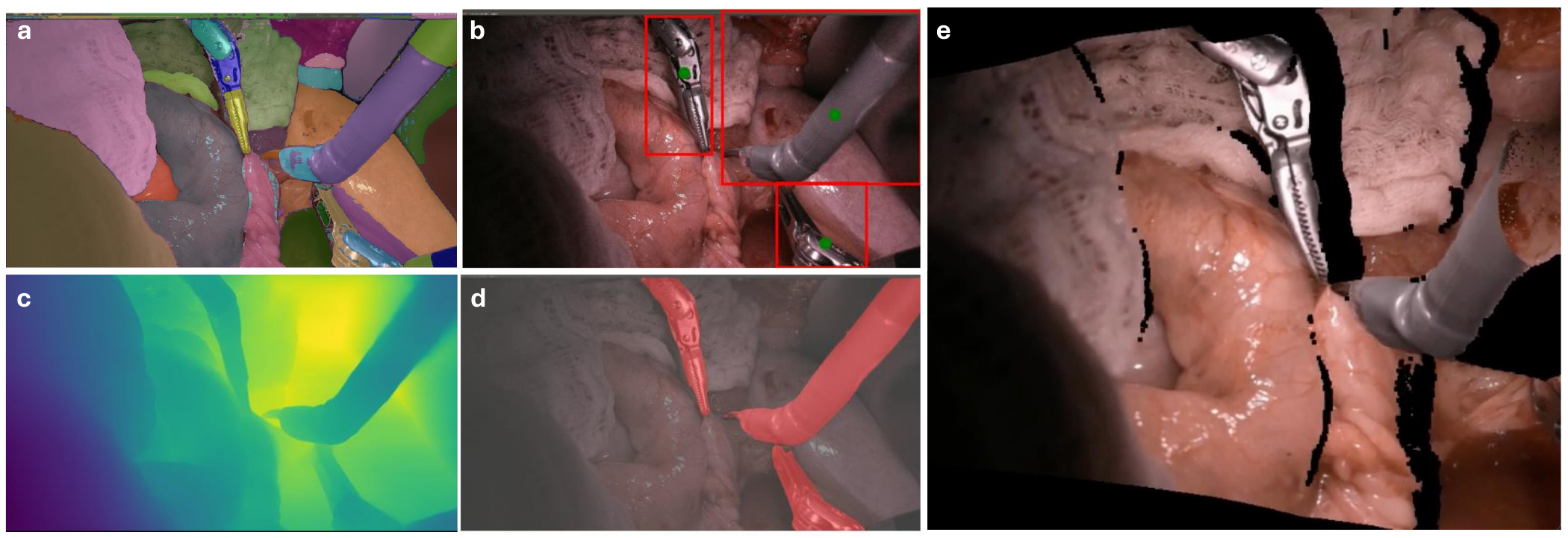}
		\caption{Examples of downstream applications: (a) instance segmentation (\cite{ravi2024sam}), (b) surgical tool detection with bounding boxes (\cite{liu2025visual}), (c) depth estimation (\cite{bochkovskii2024depth}), (d) tool segmentation ~\cite{ravi2024sam}, and (e) 3D reconstruction. (\cite{hu2025depthcrafter})}
        \label{fig: applications}
\end{figure*}
\subsection{Potential Downstream Applications Benefited by SurgiSR4K}

\subsubsection{Monoscopic Depth Estimation} Native 4K frames retain subtle photometric cues such as specular highlights, sub‑pixel texture flow, and shading that improve scale‑ambiguous monocular depth networks such as MiDaS. The accompanying 5s clips (30 fps) enable temporal photometric self‑supervision without requiring stereo rigs or structured light, providing a valuable benchmark for dense 3D reconstruction of flexible organs. In addition, the high-resolution depth information supports tissue tracking, enabling more accurate modeling of tissue deformation and interaction in real-time. Tissue tracking challenge \citep{schmidt2025point} and method \citep{chen2025mfst} emphasize the critical role of high-resolution depth estimation in precise tissue tracking. This is crucial to improve surgical precision and support the development of autonomous systems (\autoref{fig: applications} c).

\subsubsection{Surgical Instrument Detection}
Fine instrument shafts (with 0 to 2mm) and grasper tips often collapse into a single pixel in 1080p footage.  Thanks to its 4k resolution, our dataset exposes clear tool edges and metallic reflections, enabling detection of thin surgical instruments and pixel-level mask curation for semantic segmentation frameworks. Precise tool detection and localization are critical for downstream manipulation tasks such as suturing and tissue manipulations. The video nature of \textit{SurgiSR4K} also allows researchers to go beyond frame-wise detection to \emph{multi-instance tracking (\cite{li2023tatoo})}.  Accurate 3D trajectories allow quantitative workflow analysis, objective skill assessment, context-aware autonomy, and safety monitoring in minimally invasive surgery ((\autoref{fig: applications} b).

\subsubsection{Instance Segmentation} 
Fine-grained instance segmentation in surgical scenes is particularly challenging due to occlusions, specular highlights, and visual similarity between instruments and surrounding tissue. With its native 4K resolution, \textit{SurgiSR4K} allows accurate delineation of individual tool instances, even in overlapping or cluttered scenarios, by preserving fine spatial boundaries and textural cues. This enables the development of robust instance-aware models that can distinguish between multiple instruments of the same type. High-resolution instance masks also support pixel-level supervision for training advanced segmentation architectures, including transformer-based or graph-based models. Instance-level understanding is essential for precise robotic control, tool usage analysis, and temporal reasoning across frames, facilitating intelligent surgical assistance and real-time decision support (\autoref{fig: applications} a).

\subsubsection{Scene Reconstruction}
High-resolution 4K video data is fundamental for accurate 3D scene reconstruction in surgical environments, where precise spatial understanding is critical for safe navigation and intervention planning. The native 4K resolution in SurgiSR4K preserves fine geometric details such as tissue surface topology, instrument-tissue contact points, and subtle anatomical landmarks that are essential for situational understanding. Unlike lower-resolution datasets, where fine details are lost to aliasing and compression artifacts, 4K footage retains submillimeter features necessary for dense point cloud generation and mesh reconstruction. The temporal consistency provided by 30 fps video sequences enables sophisticated SLAM (Simultaneous Localization and Mapping) approaches that can track camera motion while simultaneously building detailed 3D models of the surgical field. Furthermore, high-resolution depth cues—including specular highlights, texture gradients, and shading variations—support more accurate photometric stereo techniques and 3D Guassian Splatting training for novel view synthesis. 
This enhanced 3D understanding directly benefits downstream applications such as augmented reality surgical guidance (\cite{shu2023twin}, \cite{zhang2023feasibility}), preoperative planning overlay registration, and autonomous instrument navigation in complex anatomical environments (\autoref{fig: applications} e).

\subsection{limitations}
Traditional fully laparoscopic methods (closed-cavity) were not feasible due to the size of the external camera and the limited space within the pig’s abdomen. Instead, a large incision (open abdominal cavity) was used to insert the camera.

This open approach introduced certain constraints: the camera had to remain in the chest / abdominal opening, with restricted maneuverability compared to an endoscope inside a trocar. Consequently, the freedom to adjust the camera field of view is limited, and the surgical scene is sometimes occluded due to the line-of-sight issues caused by instruments or tissue folds.

The use of an off-the-shelf 4k camera provides optics larger than those in conventional endoscopes. Although the larger aperture size increases the diffraction limit and enables high optical resolution, it has constrained the depth of field (DOF) compared to endoscopes. Frames with large out-of-focus areas have been removed during quality control. For future work, we plan to include scenes captured using large-DoF cameras.

\acks{This work was supported by Intuitive Surgical, Inc. We appreciate the valuable discussions with Max Allan and Xingtong Liu, and the support and encouragement from Kelvin (Fei) Chu, Qunming Peng, Ran Sendrovitz and Jonathan Halderman.}

%

\coi{We declare we don't have conflicts of interest.}


\bibliography{sample}


\clearpage
\appendix

\end{document}